\documentclass[journal,onecolumn,12pt]{IEEEtran}
\usepackage{graphicx}
\usepackage{booktabs}
\usepackage{makecell}
\usepackage{todonotes}
\usepackage{hyperref}
\usepackage{caption}
\usepackage{subcaption}
\ifCLASSINFOpdf
\else
\fi

\usepackage[normalem]{ulem}
\useunder{\uline}{\ul}{}
\newcommand{\nop}[1]{}

\newcommand{\eg}{\emph{e.g., }}
\newcommand{\ie}{\emph{i.e., }}

\usepackage[strict]{changepage}
\usepackage{framed}
\definecolor{formalshade}{rgb}{0.95,0.95,1}
\definecolor{darkblue}{rgb}{0.14,0.22,0.62}
\newenvironment{formal}{%
  \MakeFramed{\advance\hsize-\width\FrameRestore}%
  \noindent\hspace{-4.55pt}
  \begin{adjustwidth}{}{7pt}%
}
{%
  \end{adjustwidth}\endMakeFramed%
}

\begin{document}
%
\title{Critical Path Prioritization Dashboard for Alert-driven Attack Graphs}
%
%
%

\author{Sònia Leal Díaz*, Sergio Pastrana†, Azqa Nadeem‡\thanks{* e-mail: sonialeal01011@gmail.com}\\ %
     \scriptsize Universidad Carlos III de Madrid * † %
\thanks{† e-mail: spastran@inf.uc3m.es}\\ %
     \scriptsize Delft University of Technology ‡  
\thanks{‡ e-mail: azqa.nadeem@tudelft.nl}
}

\maketitle

\begin{abstract}
Although intrusion alerts can provide threat intelligence regarding attacker strategies, extracting such intelligence via existing tools is expensive and time-consuming. Earlier work has proposed SAGE, which generates attack graphs from intrusion alerts using unsupervised sequential machine learning.
This paper proposes a querying and prioritization-enabled visual analytics dashboard for SAGE. The dashboard has three main components: (i) a `Graph Explorer' that presents a global view of all attacker strategies, (ii) a `Timeline Viewer` that correlates attacker actions chronologically, and (iii) a `Recommender Matrix' that highlights prevalent critical alerts via a MITRE ATT\&CK-inspired attack stage matrix. 
We describe the utility of the proposed dashboard using intrusion alerts collected from a distributed multi-stage team-based attack scenario.
We evaluate the utility of the dashboard through a user study. 
Based on the responses of a small set of security practitioners, we find that the dashboard is useful in depicting attacker strategies and attack progression, but can be improved in terms of usability.  
\end{abstract}


%
\IEEEpeerreviewmaketitle

\section{Introduction}
%
%
%
%
\IEEEPARstart{S}{ecurity} 
Operations Centers (SOCs) are responsible for managing, triaging and analyzing large volumes of intrusion alerts. The processes associated with the forensic analysis of security incidents, such as determining attacker objectives and their strategies, are often manual and time-intensive. The increased workload is often deemed responsible for `alert fatigue' -- a phenomenon that leads to an increase in oversights made by security analysts, including missing critical events. Extensive research has been conducted to address `alert fatigue' \cite{app13116610} \cite{Ban_Samuel_Takahashi_Inoue_2021} \cite{10.1007/978-3-030-36708-4_62}. These works typically introduce machine learning-based visual analytics tools within analyst workflows that help them visualize large datasets, identify unique patterns, and provide actionable intelligence.

\begin{figure}[t]
    \centering
    \includegraphics[width=0.35\linewidth]{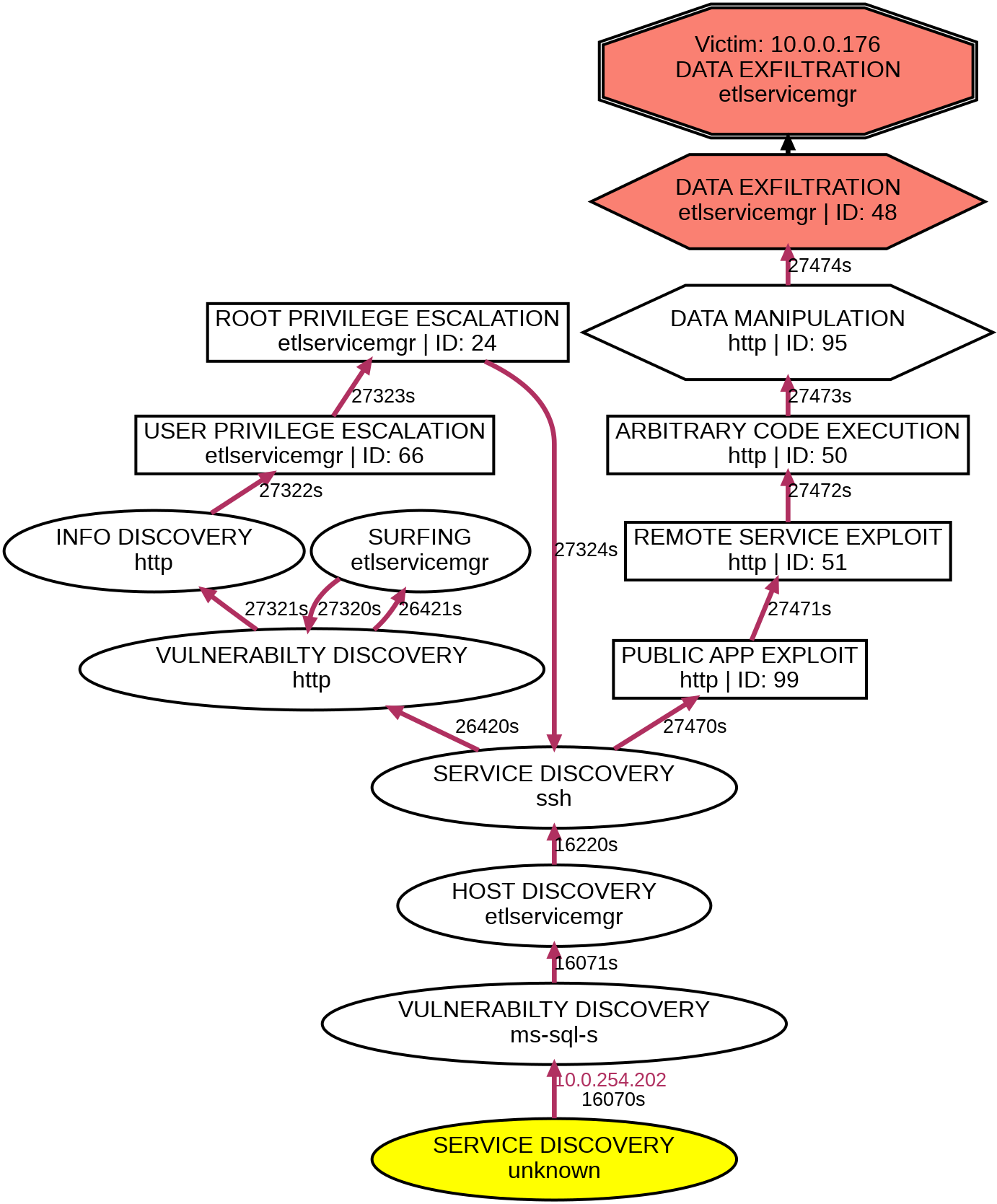}
    \caption{A SAGE attack graph: It shows one attack path to conduct data exfiltration on a victim host using the etlservicemgr service. A node label refers to the Micro attack stage, the targeted service, and a numeric context identifier. The edge labels show the seconds elapsed since the first alert was observed.}
    \label{example-ag}
\end{figure}

Recently, `alert-driven attack graphs' (AGs) have been proposed as a novel paradigm of attack graphs that reverse engineer attacker strategies from the actions observed through intrusion alerts \cite{nadeem2021alert}. 
The tool responsible for constructing these attack graphs, called SAGE, is an interpretable, unsupervised, sequential learning pipeline that compresses thousands of intrusion alerts into AGs without a priori expert knowledge about existing vulnerabilities and network topology.  
Individualized AGs are generated for every attack objective exploited on the victim host(s).  %
Each AG shows the strategies employed by all attackers against one victim host, which is then rendered as a PNG/SVG image, as shown in Figure \ref{example-ag}. 


While SAGE does the heavy lifting in terms of discovering and displaying attacker strategies from raw alerts, 
it is infeasible and time-consuming for security analysts to visualize each AG separately for finding global patterns. 
SAGE does not prioritize critical attack paths\footnote{A critical path is a series of actions that lead to critical event(s). It reflects an intrusion alert sequence that ends with a severe alert.} that might need the urgent attention of security analysts. The AG nodes are not linked back to the intrusion alert signatures, omitting crucial details for understanding the vulnerabilities exploited by the attackers. 
Additionally, while the average complexity of SAGE AGs is lower than alternative modeling approaches \cite{ccs}, the AGs of relatively common objectives (\eg data exfiltration on HTTP) can be significantly more complex with hundreds of nodes. 
The AGs do not provide any interactive component to navigate them, \ie attack paths cannot be filtered by time or attack stage of interest. Preliminary interviews with security analysts revealed that such large graphs with no interaction capabilities are unhelpful in an operational setting.

\begin{figure}[t]
  \centering
  \includegraphics[width=0.95\linewidth]{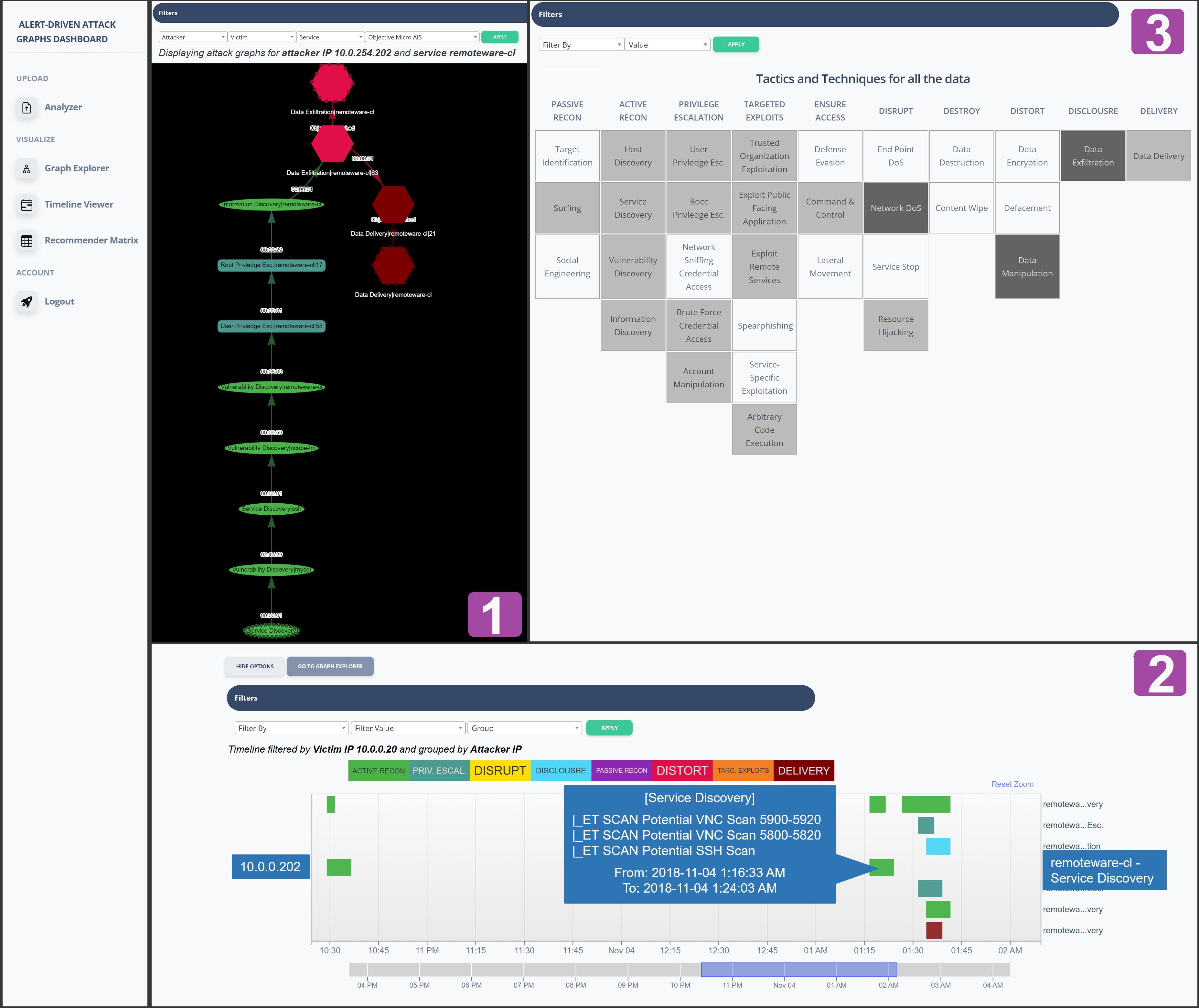}
  \caption{The critical path prioritization dashboard for alert-driven attack graphs. Attacker strategies extracted from the uploaded intrusion alerts are visualized as a: (1) \textit{Graph Explorer} showing a unified view of all strategies, (2) \textit{Timeline Viewer} enabling the analysis of individual attacker actions, and (3) \textit{Recommender Matrix} highlighting the alerts that require the urgent attention of an analyst.}
  \label{fig:teaser}
\end{figure}

In this paper, we develop a web-based visual analytics dashboard for alert-driven attack graphs with querying and prioritization capabilities for critical attack path exploration. The dashboard reduces analyst workload by i) discovering and extracting attacker strategies from intrusion alerts (done by the SAGE module), ii) consolidating the discovered attacker strategies in a dashboard with filtering capabilities, and iii) prioritizing critical events that might require an analyst's urgent attention. Figure \ref{fig:teaser} shows the three visualizations of the proposed dashboard: 
(1) \textit{Graph Explorer} presents a consolidated view of all attacker strategies discovered by SAGE, and interactively visualizes them to find relationships between different objectives.
(2) \textit{Timeline Viewer} enables the analyst to investigate temporal correlations between attacker actions at different time stamps. 
(3) \textit{Recommender Matrix} shows a condensed version of the MITRE ATT\&CK tactics/techniques in a matrix that assists the analyst in prioritizing critical events based on their prevalence and urgency in the alert dataset. 

We present the usefulness of the dashboard using intrusion alerts collected from a distributed multi-stage team-based attack scenario, captured by a Collegiate Penetration Testing Competition
(CPTC). This paper demonstrates how the dashboard enables analysts to gain insights regarding distinct paths followed by attackers and enables them to distinguish between different types of attacker strategies. We also present a preliminary empirical study by conducting a virtual survey with a small set of security practitioners. The survey asks the participants to solve various security tasks with the help of the dashboard. While the participants report that the dashboard improves their situational awareness regarding attacker strategies without time-consuming alert analysis, they also report usability issues pertaining to the dashboard's interactive components. Thus, familiarity with an interactive dashboard is a crucial indicator of its usability.

In summary, our key contributions are as follows:
\begin{enumerate}
    \item We develop a dashboard for alert-driven attack graphs with querying and prioritization capabilities for critical attack paths;
    \item We operationalize the dashboard on a distributed multi-stage team-based attack scenario, and showcase the actionable insights that can be obtained by security analysts; 
    \item We evaluate the usefulness of the proposed dashboard by conducting an empirical study with a small set of security practitioners. 
\end{enumerate}

\nop{\begin{figure}[ht]
    \centering
    \includegraphics[width=0.6\linewidth]{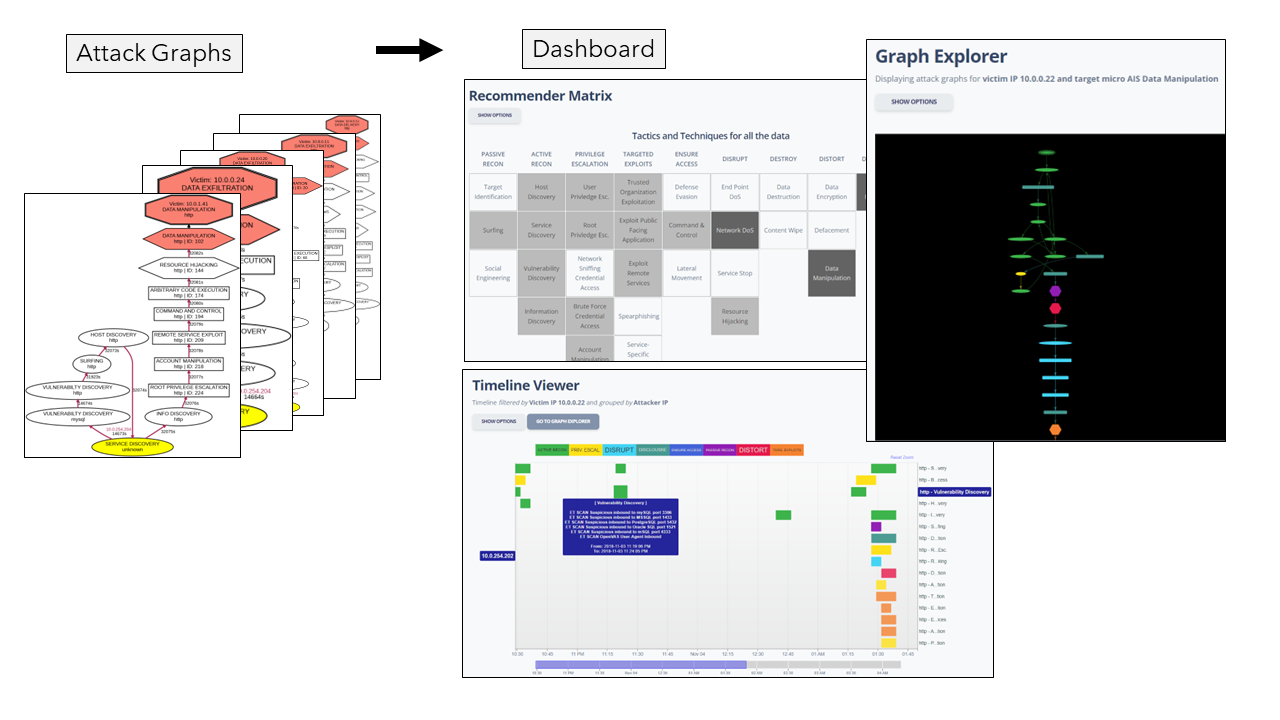}
    \caption{SAGE's attack graphs are unified into the interactive dashboard with three different views for an enhanced user experience.}
    \label{tab:AG-vs-dashboard}
\end{figure}}


\section{Related Work and Preliminaries}
Intrusion alerts are observable artifacts that can be used to reverse engineer attacker actions and their strategies \cite{moskal2020framework}.
%
Several intrusion alert-based visual analytics dashboards have been proposed in the literature. 
CyGraph \cite{NOEL2016117} is an interactive tool developed by MITRE that links intrusion alerts to vulnerability pathways. 
BubbleNet \cite{McKenna_Staheli_Fulcher_Meyer_2016} visualizes patterns extracted from network data in a Dorling cartogram, a temporal heatmap, and a bar chart.  
CRUSOE \cite{HUSAK2022102609} provides situational awareness through node-link diagrams and log tables that quantify the effects of the discovered vulnerabilities. 
IntruDTS \cite{IntruDTS} utilizes machine learning to extract patterns from network traffic logs, which are visualized using time series charts and hierarchy-host diagrams.
NViZ \cite{NViZ} visualizes the type distribution of Suricata alerts in a pie chart, employs a node-link diagram to show interactions between hosts, and provides a ``Time Controller'' slider to limit the alerts that are visualized. 

Attack graphs are a popular visualization technology to show the pathways exploited by an adversary to achieve their objectives \cite{research_on_AG_trends}. Existing literature utilizes prior knowledge about vulnerabilities \cite{ag-nessuss} and network topology \cite{ag_mtd} to construct attack graphs.
MAD \cite{MAD} creates dynamically evolving attack graphs as new vulnerabilities are discovered. In these attack graphs, the nodes are network devices, and edges illustrate CVE identifiers.
PERCIVAL \cite{percival} is another attack graph visualization tool that supports both proactive and reactive analysis of attack progression. The nodes here are network devices, while the edges illustrate the logical connections between nodes.
Early work on visualizing attack graphs such as GARNET \cite{GARNET} emphasizes network reachability and attacker simulation. It provides a ``Network Map'' to visualize the topology of the network, and ``Summary Plots" for vulnerability enumeration. 
NAVIAGATOR \cite{NAVIGATOR} aims to improve GARNET \cite{GARNET} with a more granular network asset visualization. It also facilitates the creation of hypothetical scenarios for analysts to take proactive actions.

\textbf{Alert-driven Attack Graphs.} In recent years, a new paradigm of attack graphs has been proposed that display attacker strategies learned directly from intrusion alerts without any prior expert knowledge, \ie SAGE \cite{nadeem2021alert}. 
SAGE relies on a condensed version of the MITRE ATT\&CK framework \cite{mitreattack}, called the Action-Intent Framework (AIF) \cite{moskal2020framework}, that maps intrusion alert signatures to attack stages, \ie Action Intent States (AIS). The Micro AIS denote the techniques, and the Macro AIS aggregate Micro AIS to reflect the high-level tactics.
%
Bursts of intrusion alerts with the same attack stage are aggregated into `episode sequences' (reflecting a series of attacker actions) for each attacker/victim host pair. A suffix-based probabilistic deterministic finite automaton (S-PDFA) is learned from the episode sequences that model the context in which episodes (attacker actions) appear, and the outcome they lead to. The episode sequences, together with their modeled context, are transformed into alert-driven attack graphs for each objective exploited on victim host(s). 
A single attack graph generated by SAGE shows how an attack transpired on a victim host, and all the attacker hosts that reached this objective. The context modeled by the S-PDFA (shown by a so-called \textit{context identifier}) is used to identify significantly different ways of reaching the objective. Thus, an analyst can visualize attack graphs of interest in order to obtain threat intelligence regarding similar strategies, potentially scripted attempts, attacker behavior dynamics, and fingerprintable paths. 



\section{Critical Path Prioritization Dashboard}

\begin{figure*}[t]
    \centering
\includegraphics[width=0.9\linewidth]{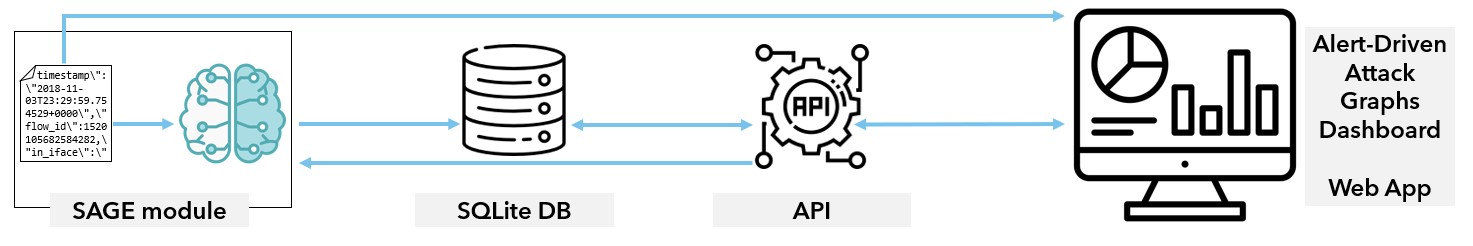}
  \caption{Dashboard workflow: The API triggers the execution of SAGE, which parses the alerts and learns the attack graphs. The API stores them in an SQLite database, and populates the web interface.}
  \label{fig:workflow}
\end{figure*}

Although SAGE drastically reduces the volume of alerts to be analyzed (by 99.9\% \cite{nadeem2021alert}), the attack graphs still need to be individually investigated by a security analyst. 
This section describes the proposed web-based visual analytics dashboard, which provides a consolidated view of the alert-driven attack graphs generated by SAGE. 
%
%
The dashboard enables analysts to extract actionable insights regarding attacker strategies, such as the paths followed by attackers to exploit specific vulnerabilities, the attack progression over time, and the most urgent technique observed. 
As depicted in Figure \ref{fig:workflow}, there are three modules of the dashboard: (i) the web interface, (ii) the back-end database, and (iii) the RESTful API. 
\begin{enumerate}
    \item \textbf{Web interface:} The web interface is composed of the three visualizations, \ie Graph Explorer, Timeline Viewer, and Recommender Matrix. It uses an AppSeed\footnote{\url{https://appseed.us/}} template, which is a platform that provides templates for the automated creation of modern, full-stack projects. We selected the Django Soft UI Dashboard  template, created by Creative Tim. It uses Django, HTML5, Twig, and Bootstrap 5 in the starter kit. The dashboard has a menu located on the left-hand side with the three visualization options. Each visualization contains filters and additional information regarding, \eg actions that can be taken on the current view, and interpretation of the respective view elements. 
    \item \textbf{Database:} We use a relational SQLite database to store the attack graphs generated by SAGE. We preserve the contextual information while storing attack graph nodes in the database. Note that the Graph Explorer de-duplicates nodes that share the same Micro AIS, targeted service, and context identifier, in order to provide a holistic view of the attack paths. 
    \item \textbf{API:} The Flask-based RESTful API is an intermediary between the database and the Django web application. The RESTful API is responsible for two main actions: (a) Retrieve the requested attack graphs from the database in order to populate the various views of the dashboard, and (b) Update the severity weights, severity levels and urgency ranges set by the user in the Recommender Matrix. In addition, the API offers the necessary data needed for automated report generation. However, this has been left as future work. 
\end{enumerate}

\subsection{Dashboard Architecture}

Intrusion alert files can be uploaded in JSON format to the dashboard. This triggers the execution of the SAGE module, which processes the alerts, applies unsupervised sequence learning, extracts temporal and probabilistic patterns from the alerts, and produces state sequences that reflect attacker strategies. 
The distinct visualizations emphasize different aspects of the attacks: (1) The Graph Explorer shows a global, unified view of all attacker strategies. (2) The Timeline Viewer illustrates the tactics/techniques used by the attackers at different time stamps, \eg to determine if an attacker targets numerous victims at the same time. Note that the Graph Explorer shows attack paths, while the Timeline Viewer provides a deeper temporal view of the attacker actions. (3) The Recommender Matrix shows a condensed version of the MITRE ATT\&CK framework where the different techniques are highlighted based on the prevalence and urgency of critical events. The analyst can click on one of the highlighted attack stages to be redirected to the Graph Explorer, which shows only the attack graphs containing this technique. 



%

\subsubsection{\textbf{Graph Explorer}}
The Graph Explorer is implemented using the ``Network'' feature of \textit{vis.js} --- a JavaScript library that enables efficient interactions with large graphs.
The Graph Explorer consolidates all alert-driven attack graphs generated by SAGE. Combining all attack graphs within the same view enables an analyst to draw conclusions regarding the most frequently used pathways towards an objective, and the temporal inter-dependency between the various objectives. Filters can be used to focus on specific attack paths based on the attacker host, victim host, targeted service, and attack stage (see Figure \ref{fig:teaser}-1). 

\begin{figure}[t]
\centering
\begin{subfigure}{.7\textwidth}
  \centering
   \includegraphics[width=\linewidth]{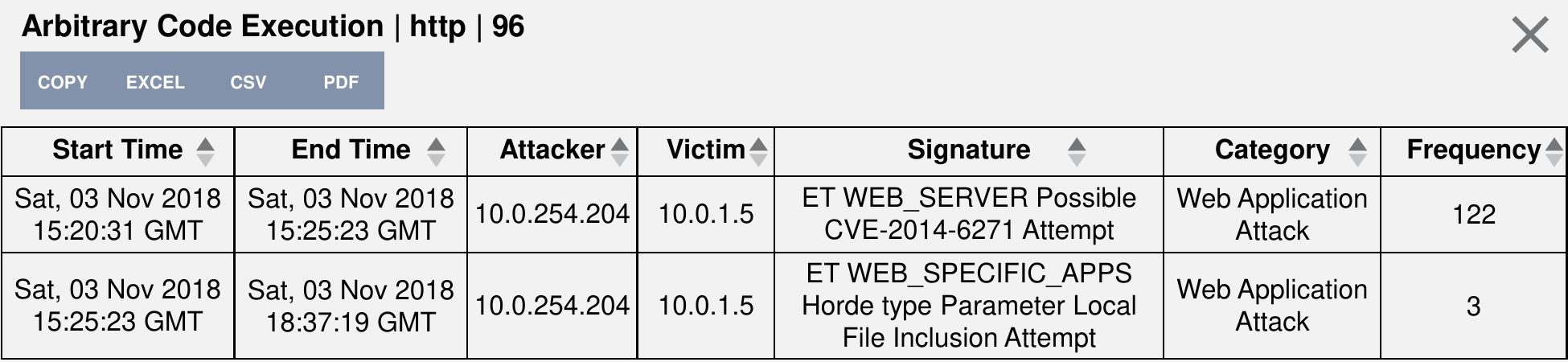}
    \caption{The signature table is displayed at the top of the dashboard when a node is clicked in the Graph Explorer. It shows various alert attributes that can be sorted, if needed. The table can be exported in various formats for automated report generation.}
    \label{tab:ge-table}
\end{subfigure}\hfill
\begin{subfigure}{.25\textwidth}
  \centering
  \includegraphics[width=\linewidth]{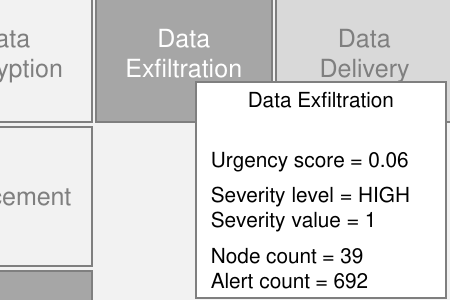}
    \caption{A tool-tip for the Recommender Matrix cell shows the computed urgency score, the severity level/weight, and the number of alerts and attack graph nodes for the Micro AIS.}
    \label{fig:details-rm}
\end{subfigure}
\caption{Dashboard elements: (a) Signature table; (b) Urgency tool-tip.}
\label{fig:test}
\end{figure}

The nodes and edges in the graph represent similar information as the attack graphs from SAGE: \textit{Nodes} represent attacker actions (Micro attack stage, targeted service, context identifier); \textit{Node shape} reflects the severity of the node (ellipses are low-severity, boxes are medium-severity, and hexagons are high-severity exploited objectives), and \textit{Edge labels} show the time elapsed between attacker actions in the format HH:MM:SS. 
The graph encodes the following additional information: (a) \textit{Node color} reflects the type of Macro AIS (tactic). (b) \textit{Node Border} represents the start and end of an attack path (a dotted-ellipse represents the start, and a dotted-hexagon represents the end). 
Note that all objectives converge to a single artificial node with an unspecified context identifier to have a common root node.
(c) A \textit{single-click} on a node displays a signature table that shows the alert signatures that are triggered by the corresponding attacker action (see Figure \ref{tab:ge-table}). It also shows the start and end time of the attacker action (episode), the attacker and victim IP addresses, and the frequency of the different alert signatures. This provides crucial information regarding the techniques employed by the attackers, and is also useful for differentiating between attack variants. 

The layout of the graph is hierarchical, with two supported sorting methods: Directed and Hubsize. \textit{Directed} sorts the nodes based on their direction from source to destination. \textit{Hubsize} (default method) sorts the nodes based on their edge frequency. Depending on the attack graph at hand, one sorting method might be more suitable than the other. For instance, the \textit{Directed} method allows analyzing the flow of attacks and tracing attack progression, while the \textit{Hubsize} method can be advantageous when identifying junction nodes and uncovering critical dependencies between attack stages.



\subsubsection{\textbf{Timeline Viewer}}
The Timeline Viewer is implemented using the Timelines Chart (based on D3 timelines-chart), which supports \textit{swimlanes} to describe time-series data. 
The Timeline Viewer enables a user to focus on specific attacker actions that occurred during a user-selected time window. Essentially, the nodes in the Graph Explorer are shown as colored segments in the Timeline Viewer (see Figure \ref{fig:teaser}-2). The \textit{segment color} represents the corresponding Macro AIS (same as in the Graph Explorer).
Note that we specifically chose two different views to investigate attacker actions and attack paths to reduce the cognitive load on the analysts.
This view can be used to determine, \eg if an attacker is targeting numerous victims at the same time.
Depending on the time progression between segments, analysts can even infer attacker tooling, \eg several simultaneous actions may indicate the usage of automated vulnerability scanners.

The swimlanes can be configured to either show the attacker perspective or the victim perspective, \ie the swimlanes only show segments corresponding to the chosen attacker or victim host. Within a swimlane, there are numerous rows that have different \textit{labels}. The labels correspond to a combination of the Micro AIS and targeted service, which enable displaying multiple techniques per service. Each segment supports a \textit{tool-tip}, which displays alert signatures (similar to the Graph Explorer). There is a \textit{timeline brush} located at the bottom to zoom into specific time ranges. Dragging into a specific region is also possible.  
Finally, a \textit{``Go to Graph Explorer'' button} allows visualizing the current attack stages displayed in the Timeline Viewer as attack paths in the Graph Explorer. It takes the information currently shown (\ie attacker and victim IP, start and end times, and targeted service), and displays only those attack paths that meet the defined criteria.

\subsubsection{\textbf{Recommender Matrix}}

The Recommender Matrix is inspired by MITRE ATT\&CK Navigator\footnote{\url{https://mitre-attack.github.io/attack-navigator/}}. It assists analysts in prioritizing critical attacker actions based on their urgency. 

The urgency score for each Micro AIS is calculated based on two factors --- severity and prevalence (see Eq. \ref{eq:urgency}). \textit{Severity} reflects the level of criticality of an attack stage. \textit{Prevalence} refers to the occurrence frequency of the attack stage. We \textit{normalize the prevalence} of Micro AIS to accommodate for the imbalance in the alert types, shown by Eq. \ref{eq:prev}. Here, \textit{N} is the set of all attack graph nodes, and \textit{Count(N,Micro)} gives the count of nodes that have the specific Micro AIS. 
Each Micro AIS is assigned a default \textit{severity level} according to the original SAGE paper \cite{nadeem2021alert}. We assign \textit{severity weights} to these severity levels, \ie low-severity (0.25), medium-severity (0.5), and high-severity (1.0). The dashboard allows analysts to modify the severity levels for each attack stage, and the weights of the severity levels based on the risk appetite of their SOC. For example, a safety-critical organization might consider `Host Discovery' alerts as severe, and modify their severity level from `Low' to `Medium', and may even assign a higher severity weight to them. 
The urgency score is thus a trade-off between the severity and frequency of alerts. For instance, there might be cases where techniques occur less often, but require immediate attention since they are highly severe. The aim of the urgency score is to create a balance between severe but infrequent alerts (\eg data exfiltration) vs. non-severe but frequent alerts (\eg host discovery). 

\begin{equation}
    urgency\_score (Micro) = severity\_weight (Micro) * normalized\_prevalence (Micro)
    \label{eq:urgency}
\end{equation}

\begin{equation}
normalized\_prevalence (Micro) = \frac{Count(N,Micro)}{|N|} 
    \label{eq:prev}
\end{equation}

The Recommender Matrix shows the following information: The cells in the \textit{top row} of the matrix show Macro AIS. For each Macro AIS, its corresponding Micro AIS are located in the cells below it  (see Figure \ref{fig:teaser}-3). The \textit{cell color} is based on the computed urgency score. We support three gray-scale colors, referring to minor (light gray), major (medium gray), and critical (dark gray) events. The thresholds for the color ranges can be customized by the SOC analysts. More granular urgency scores can be computed for specific victim IP, attacker IP, or targeted service using provided filters. 
Each cell supports a \textit{tool-tip}, which shows the urgency score, severity level, and severity weight of the corresponding Micro AIS. It also shows the frequency of alerts, and the number of attack graph nodes that correspond to the Micro AIS (see Figure \ref{fig:details-rm}). 
\textit{Clicking} on a cell with a non-zero urgency score redirects the user to the Graph Explorer, which shows the specific attack paths where the particular Micro AIS occurs (which is \textit{colored in white} for easier identification). This allows the analysts to gain a broader perspective on the different kinds of attacks that were enabled due to the selected technique.  

%

\begin{figure}[t]
    \centering
    
\end{figure}


\section{Experimental Dataset and Exemplary Use Case}

\subsection{Experimental Dataset}\label{sec:dataset}
We populate the dashboard using intrusion alerts collected from a distributed multi-stage team-based attack scenario collected during a Collegiate Penetration Testing Competition (CPTC) in 2018\footnote{\url{https://mirror.rit.edu/cptc/2018/}}. The dataset contains  330,270 Suricata alerts generated by 6 student teams over 9 hours. For the purposes of this study, we utilize the alerts generated by Team 1 (resulting in 81,373 alerts). SAGE compresses the alerts into 53 AGs, while the unified dashboard allows the user study participants to explore and reason about the discovered attacker strategies.

\subsection{Exemplary Use Case}
The Recommender Matrix shows that alerts associated with \textit{Data Exfiltration}, \textit{Data Manipulation}, and \textit{Network DoS} require the urgent attention of security analysts. 

\par{\textbf{Data exfiltration attempts.}} A security analyst can view the most common strategies used by attackers to exfiltrate data over HTTP by using the \textit{Data Exfiltration} and \textit{http} filters in the Graph Explorer. All of the filtered pathways perform \textit{Root Privileged Escalation} and \textit{Data Manipulation} before exfiltration. By clicking on one of the privilege escalation nodes, the alert signature table shows, \eg ``GPL EXPLOIT CodeRed v2 root.exe access'' and ``ET WEB\_SERVER ColdFusion administrator access'', providing actionable intelligence that the CodeRED and ColdFusion exploits were used to carry out this attack. 

\par{\textbf{Attacks after working hours.}} A security analyst may seek to investigate if any anomalous activity occurred at specific times, such as after working hours or on holidays. The Timeline Viewer can be used to investigate what transpired on a victim host. For example, it shows that the victim 10.0.0.22 was targeted by the attacker 10.0.254.202 until 01:40 AM.
The analyst can view the strategies employed during this time by redirecting to the Graph Explorer, which illustrates that \textit{Root Privilege Escalation} enabled \textit{Resource Hijacking}, which further enabled \textit{Data Exfiltration}, \textit{Arbitrary Code Execution}, and \textit{Data Manipulation}. The analyst can use this information to determine which resources were compromised during these attacks.

\begin{table}[t]
\begin{minipage}{.5\linewidth}
    \centering

    \caption{Survey participants' demographics}
\label{tab:demographics}

    \medskip

\begin{tabular}{ccc}
\toprule
\textbf{Participant} 
& \textbf{Job title} 
& \textbf{Experience}  \\ \midrule
P1 & Information Security Officer & 1-5 years \\ 
P2 & Incident Handler & 1-5 years \\ 
P3 & Senior Security Consultant & \textgreater 5 years \\ \bottomrule
\end{tabular}
\end{minipage}\hfill
\begin{minipage}{.5\linewidth}
    \centering

    \caption{Number of participants per user study task}
\label{tab:participants}

    \medskip

\begin{tabular}{cc}
\toprule
\textbf{Tasks}                  & \textbf{Participants} \\ \midrule
Task 1                         & 3                               \\ 
Task 2                         & 2                               \\ 
Task 3                         & 1                               \\ 
User Experience & 1                               \\ \bottomrule
\end{tabular}
\end{minipage}

\end{table}

\begin{table}[t]
\caption{Survey statements and the average responses on a 5-point Likert scale (1/5=Completely disagree/agree).}
\label{tab:survey-responses}
\resizebox{\columnwidth}{!}{%
\begin{tabular}{lc}
\toprule
\multicolumn{1}{c}{\textsc{\textbf{Statement}}} & \textsc{\textbf{Average score}} \\ \midrule
\multicolumn{2}{c}{\textbf{Task 1}} \\ \midrule
\multicolumn{1}{l}{The Recommender Matrix reduces the time to identify critical techniques/objectives as opposed to the manual analysis of alerts.} & 3.67 \\ 
\multicolumn{1}{l}{The colors used in the matrix convey the urgency of each technique/objective.} & 2.67 \\ 
\multicolumn{1}{l}{It is easy to identify the most commonly used attacker techniques/objectives that should be given priority in my job.} & 3.33 \\ 
\multicolumn{1}{l}{The Recommender Matrix is user-friendly.} & 3.33 \\ \midrule
\multicolumn{2}{c}{\textbf{Task 2}} \\ \midrule
\multicolumn{1}{l}{The Graph Explorer makes it easy to see the attack progression.} & 4.5 \\ 
\multicolumn{1}{l}{The Graph Explorer reduces the time it takes to identify attack campaigns compared to the alert tools I currently use.} & 4 \\ 
\multicolumn{1}{l}{The Graph Explorer filtering capability makes it easy to pinpoint vulnerable services on the victim hosts.} & 4.5 \\ 
\multicolumn{1}{l}{The Graph Explorer makes it easier to discover relationships between different attack stages.} & 5 \\ 
\multicolumn{1}{l}{The table with alert signatures helps me identify the procedures carried out by an attacker.} & 3 \\ 
\multicolumn{1}{l}{The node colors are useful for distinguishing between different attack stages.} & 4 \\ 
\multicolumn{1}{l}{The Graph Explorer is user-friendly.} & 3 \\ \midrule
\multicolumn{2}{c}{\textbf{Task 3}} \\ \midrule
\multicolumn{1}{l}{The Timeline Viewer simplifies the comprehension of attack progression compared to the alert tools I normally use.} & 5 \\ 
\multicolumn{1}{l}{The Timeline Viewer helps to identify different tooling used by an attacker.} & 4 \\ 
\multicolumn{1}{l}{The Timeline Viewer filtering capability facilitates the analysis of attack evolution for a desired victim or attacker.} & 3 \\ 
\multicolumn{1}{l}{The Timeline Viewer is user-friendly.} & 2 \\ \midrule
\multicolumn{2}{c}{\textbf{User Experience}} \\ \midrule
\multicolumn{1}{l}{The dashboard provides situational awareness without the need for time-consuming alert analysis.} & 5 \\ 
\multicolumn{1}{l}{The dashboard helps me triage alerts.} & 4 \\ 
\multicolumn{1}{l}{The dashboard is customizable to fit my organization’s needs.} & 3 \\ 
\multicolumn{1}{l}{I would use this dashboard in my job.} & 4 \\ \bottomrule
\end{tabular}}
\end{table}

\section{User Study Operationalization}

\subsection{Qualitative User Evaluation}
We evaluated the dashboard (from Section \ref{sec:dataset}) with an empirical user study. 
We solicited security practitioners with current or prior alert management experience. The invitations were sent via email to security experts based in Spain and the Netherlands, and were also broadcasted through LinkedIn.
We obtained the necessary ethics approval from our institutional IRBs for the evaluation.
The survey was created using the GDPR-compliant tool, Qualtrics, and no personally identifiable information was collected. It was publicly available for four weeks during mid-2023.

The survey started with an information sheet where the participants were informed about the purpose of the study, and its voluntary and anonymized nature, as well as an estimated effort of 30 minutes. It also included a link to an introductory video\footnote{Introductory video of the dashboard: 
\url{https://youtu.be/9RpchlogTzs}
} so they could familiarize themselves with the dashboard. 
The survey included a hypothetical scenario for an incident occurring in a simulated environment (as part of the penetration testing competition).
The participants were then asked to complete three tasks via the dashboard, where they had to use the different visualizations to answer questions about the incident. Task 1 required the use of the Recommender Matrix, Task 2 required the use of the Graph Explorer, and Task 3 required the use of the Timeline Viewer. 
The participants were also asked to rate various aspects of each visualization on a 5-point Likert scale.
Finally, they were asked questions about the usability of the whole dashboard using free-response text boxes. 

Three security experts participated in the study, see Table \ref{tab:demographics} for their demographics. They reported spending approximately 20\%-60\% of their time per week analyzing alerts. The alert management tools they currently employ are, \eg ArcSight Enterprise Security Manager, McAfee Enterprise Security Manager, IBM QRadar, IBM Guardian, and Splunk. 
Although not all of these participants completed the survey (Table \ref{tab:participants} reports participants per task), their partial responses were nevertheless helpful for the analysis.



%

\subsection{Security Scenario}

\vspace{-7pt}
\begin{formal}
\textbf{Scenario:} \textit{The CEO of a fictitious company is being blackmailed for data exposure. In one of the company’s endpoints, whose IP address is 10.0.0.20, malicious behavior has been detected, possibly linked to the data exposure blackmail. This has raised concerns for the company.
You, as an analyst, are in charge of understanding what has happened to this victim host, and, if possible, determining the source of the incident and the sequence of steps used to conduct the attack.}
\end{formal}
\vspace{-7pt}

The primary objective of the scenario and the follow-up tasks was to construct a story that unraveled a security incident as the analysts discover attacker strategies using the dashboard.

\vspace{-7pt}
\begin{formal}
\textbf{Task 1}: \textit{Acting promptly is essential to avoid further damage, so you should focus first on the most urgent technique/objective that has targeted IP 10.0.0.20. Therefore, you must identify which technique requires immediate attention, denoted with the highest urgency score.}

\textit{Hint 1: Use the Recommender Matrix.}

\textit{Hint 2: Use the default values for `Severity weights', `Severity levels', and `Urgency ranges'.}
\end{formal}
\vspace{-7pt}

By setting a filter for the targeted victim and examining the Recommender Matrix, all three participants correctly identified that `Data Exfiltration' was the technique with the highest urgency score. P2 was able to deduce that multiple vulnerabilities had been exploited to perform malicious actions on the specified host. However, P1 noted that the first action-item of any incident is determining how a threat actor gained access, and subsequently, taking appropriate actions to remove their access. This comment draws attention to the varying definitions of `urgent technique' for different analysis strategies. In the dashboard (and according to \cite{nadeem2021alert}), the term `objective' pertains to a final high-severity action (Micro AIS) in an attack campaign, instead of the low/medium-severity action that enables attacker access. 

Having used the Recommender Matrix, the participants were asked to rate their experience on a 5-point Likert scale. Table \ref{tab:survey-responses} shows the average score for each statement, where \textit{1 reflects strong disagreement}, and \textit{5 reflects strong agreement} with the statement. 
P1 strongly disagreed with all the statements -- they did not find the visualization useful, as the actual scope of the incident could not be obtained. They stated that they preferred to see correlations between alerts rather than the attack paths, indicating that security practitioners are resistant to changes in the tool-sets that they use.
In contrast, P2 and P3 found the Recommender Matrix useful for identifying critical techniques in comparison to the tools they currently employed.
However, the colors used to convey urgency scores were not helpful according to the participants.

At this stage, P1 dropped out. P2 and P3 continued the following task.

\vspace{-7pt}
\begin{formal}
\textbf{Task 2}: \textit{Determine which services running on victim 10.0.0.20 were impacted by the previously identified technique/objective, and what are the attacker IPs targeting the victim host. In addition, has 10.0.0.20 been compromised in any other way before the previously identified technique/objective?}

\textit{Hint 1: Use the Recommender Matrix as a starting point, and then use the Graph Explorer.}

\textit{Hint 2: You can look at the alerts of a specific attack stage node by clicking on it}.
\end{formal}
\vspace{-7pt}

The participants directly went to the Graph Explorer, and filtered by the victim IP 10.0.0.20 to observe the attack paths. Note that none of the participants followed the first hint, or filter by the attack stage in the Graph Explorer. 
Both P2 and P3 correctly identified that an SQL injection was used to exfiltrate data. P2 reported that remoteware-cl and mysql services were exploited, while P3 reported that mongodb was used. The correct answer was remoteware-cl and mongodb, which could have been observed in the nodes of the Graph Explorer and the signature table: There were three `Data Exfiltration' nodes, two of which stated remoteware-cl as the targeted service. The signature table for the third node showed ``ETPRO ATTACK\_RESPONSE MongoDB Database numeration Request''. 



The IPs of the attackers were not correctly identified by the participants. While P2 selected 10.0.254.103, .202, .204, and .206, P3 only selected 10.0.254.206. The correct answer was 10.0.254.202, .204, and .206. We believe that the participants were unable to answer the question correctly since they did not filter by attack stage, and were thus seeing additional attack paths that were unrelated to the task. If they had been redirected from the Recommender Matrix, they would not have seen those extra edges. 

P2 accurately identified additional attacks on the victim host, \ie `Data Manipulation', `Network DoS', and `Data Delivery' by following their inter-dependency in the Graph Explorer and investigating the corresponding alert signatures. They described that the attack started with web scans, detecting vulnerable services that enabled the attacker to manipulate accounts, elevate privileges, and even execute arbitrary code. P3 was unable to answer this question, and dropped out of the study.

Having used the Graph Explorer, the participants were asked to rate their experience on a 5-point Likert scale (see Table \ref{tab:survey-responses}). 
Both, P2 and P3 agreed that the Graph Explorer reduced the time it took for identifying the relationships among attack stages compared to the alert management tools they currently used. The node coloring helped them distinguish between different tactics. The alert signature table was not utilized to the extent that we had anticipated. The participants also reported that they did not find the Graph Explorer to be very user-friendly because of issues pertaining to non-uniform and abrupt scrolling.

\vspace{-7pt}
\begin{formal}
\textbf{Task 3}: \textit{How did the attack on 10.0.0.20 occur using remoteware-cl? Did the attackers target the victim host at the same time? What type of tooling could the attackers have used?}

\textit{Hint 1: Use the Timeline Viewer.}

\textit{Hint 2: The services are listed on the right side.}
\end{formal}
\vspace{-7pt}

This task required the use of the Timeline Viewer for comparing how the different attackers compromised the victim host using remoteware-cl.
P2 accurately identified that the attacks were conducted during different time-windows by focusing on the area of interest of the timeline. They stated that the brief duration of the attacks might hint at the use of vulnerability scanning tools, \eg Nikto. They also identified three phases of the attack: (1) active reconnaissance, (2) passive reconnaissance and disclosure, and (3) data delivery and distortion. They were able to pinpoint the specific time-frames of the attack phases.

\nop{
The scoring of the statements is provided below:
\begin{itemize}
    \item \textit{The Timeline Viewer simplifies the comprehension of attack progression compared to the alert tools I normally use.} (5 - strongly agree)
    \item \textit{The alert tools I currently use visualize the alerts chronologically in the form of a timeline.} (4 - somewhat agree)
    \item \textit{The Timeline Viewer helps to identify different tooling used by an attacker.} (4 - somewhat agree)
    \item \textit{The filtering capabilities facilitate the analysis of attack evolution for a desired victim or attacker. }(3 - neither agree nor disagree)
    \item \textit{It is useful to visualize the attack stages in the form of a timeline.} (5 - strongly agree)
    \item \textit{The Timeline Viewer is user-friendly.} (2 - somewhat disagree)
\end{itemize}}

Based on the survey responses (see Table \ref{tab:survey-responses}), the usability of the Timeline Viewer can be improved. At the same time, chronologically organizing attack stages is useful for attack analysis, suggesting the merit of the Timeline Viewer. 

\vspace{-7pt}
\begin{formal}
\textbf{User Experience}: \textit{How was your experience using the dashboard?}
\end{formal}
\vspace{-7pt}

Finally, the participants were asked to provide their opinion on two particular aspects: usefulness of the dashboard, and areas of improvement. Table \ref{tab:survey-responses} shows the response of P2 for the Likert-scale statements.  
The participant found that the representation of the tactics/techniques was useful in the Timeline Viewer and Graph Explorer. However, there were usability obstacles in these visualizations related to scrolling. 
P2 showed interest in a module to visualize all services of a certain victim host, and suggested including a threat intelligence component to gather all the indicators of compromise detected in the attacks.

\nop{
\begin{itemize}
    \item \textit{The dashboard helps me triage alerts. }(4 - somewhat agree)
    \item \textit{The dashboard provides situational awareness without the need for time-consuming alert analysis.} (5 - strongly agree)
    \item \textit{The dashboard enables me to investigate alerts more efficiently. }(3 - neither agree nor disagree)
    \item \textit{The dashboard is customizable to fit my organization’s needs.} (3 - neither agree nor disagree)
    \item \textit{I would use this dashboard in my job.} (4 - somewhat agree)
\end{itemize}}

\section{Discussion and Future Work}

\textbf{Discussion.} Despite the low participation rate of the survey, the responses provided valuable insights into the dashboard utility. Overall, the dashboard received positive feedback in terms of providing situational awareness without having to invest significant time investigating individual alerts.
The Recommender Matrix received the highest scores in terms of user-friendliness out of all the three visualizations. One suggestion is to re-evaluate the use of colors to represent urgency, as the current gray-scale choice did not strongly convey the intended message. Nonetheless, the participants expressed that the matrix enabled them to identify the most critical techniques more efficiently compared to the alert management tools they currently used. 
Additionally, a notable strength of the Graph Explorer was its ability to illustrate the relationships among attack stages.
The participants also strongly believed that the Timeline Viewer made it easier to understand attack progression compared to the tools they currently used.

While the participants found merit in the visualizations, there were usability issues that made it difficult for the participants to appropriately adjust them. 
The Timeline Viewer suffers from significant usability issues. Based on the feedback, we believe that it provides insightful information, but the representation needs to be tweaked. 
Additionally, some features were not utilized to the extent that we were anticipating, \eg the redirection option from the Recommender Matrix to the Graph Explorer. This could be attributed to easier ways of finding the same information, lack of utility, or insufficient familiarity. This suggests the need to evaluate the user interface separately from the attacker strategy analysis --- the evaluation of the dashboard has a conditional dependency on the user interface design. In other words, a misaligned user interface interferes with the utility evaluation of the dashboard. This is why it is paramount to include security practitioners (who have an intimate understanding of their workflows) in the design phase of the dashboard that is meant to address their needs.

The user study had a low participation rate, which might have been caused by several factors. Perhaps the way in which the user study was promoted did not effectively reach a wide audience. Additionally, the duration of the user study, estimated at 30 minutes, might have discouraged people from completing the survey --- 31 people accessed the survey but only 1 fully completed it. Another factor could be the lack of incentives to complete the study in its entirety. Nevertheless, user studies with security practitioners are known to be expensive, and often suffer from low response rates \cite{nadeem2022sok}. 

\textbf{Future work.} In the future, we will investigate the use of a nested hierarchical visualization for large attack graphs. This is because one of the participants, P1, who dropped out mid-study specified that the Graph Explorer may be hard to interpret for already over-worked security practitioners. One way to improve the user experience is to create a nested-graph that can be expanded as the user delves deeper into an investigation.
Moreover, we will develop a threat intelligence module that can collect numerous indicators of compromise from heterogeneous data sources (as suggested by P2). As such, the dashboard (together with SAGE) acts like a SOAR system. While SAGE currently only supports Suricata-like intrusion alerts, additional support for network traffic, system logs and other input sources will enable analysts to corroborate an incident through different evidence sources. A key challenge here is the automated alignment of the different data sources with respect to specific incidents. Performance and scalability of the dashboard are important considerations when considering additional data sources.
Finally, we will include practitioners in the re-design phase of the dashboard to understand what features/tools they consider important to carry out incident response tasks. 

\section{Conclusions}
We propose a web-based dashboard for the alert-driven attack graphs generated by SAGE. 
It addresses the limitations of SAGE AGs in the following ways: 
The \textit{Recommender Matrix} highlights critical attacker actions that require the urgent attention of analysts. Based on the unique circumstances of a SOC, analysts can adjust the threshold for what is considered \textit{urgent}.
The proposed dashboard provides a unified view of all the attack graphs in a \textit{Graph Explorer}, also showing the inter-dependency between the various `objectives'. 
The \textit{Timeline Viewer} allows an analyst to filter attacker actions based on a selected time window.
The Graph Explorer and the Timeline Viewer also show alert signatures corresponding to attacker actions. 
Finally, in order to reduce the cognitive load on analysts, the proposed dashboard is equipped with filtering and querying capabilities based on various criteria. 

We evaluate the dashboard (populated with alerts collected through a penetrating testing competition) with a small set of security practitioners. The feedback obtained through the empirical user study demonstrates that the dashboard provides valuable insights into attacker strategies and saves analyst time.
It showcases the utility of the dashboard compared to the currently employed alert management tools, but also highlights usability issues in the user interface. This finding points to a larger gap between industry and academia -- direct collaboration with security analysts is essential to ensure that the security dashboards do not have a steep learning curve, and do not discourage analysts from adopting new tools.

\section*{Acknowledgment}
The authors wish to thank all the security practitioners who participated in the user study for their valuable insights and time. Their involvement was crucial to conduct our research.
This work has been partially supported by Grant TED2021-132170A-I00 funded by MCIN/AEI/ 10.13039/501100011033 and by the “European Union NextGenerationEU/PRTR”.

\ifCLASSOPTIONcaptionsoff
  \newpage
\fi



%
\bibliographystyle{IEEEtran}
\bibliography{bibliography}
\end{document}